# "To wet or not to wet: that is the question"

Silvina M. Gatica[a] and Milton W. Cole[b]

[a]Department of Physics and Astronomy, Howard University, 2355 Sixth Street, NW, Washington, DC 20059

[b]Department of Physics, Pennsylvania State University, University Park, PA 16802

[a]Communicating author: sgatica@howard.edu

## Abstract

Wetting transitions have been predicted and observed to occur for various combinations of fluids and surfaces. This paper describes the origin of such transitions, for liquid films on solid surfaces, in terms of the gas-surface interaction potentials V($\mathbf{r}$), which depend on the specific adsorption system. The transitions of light inert gases and $H_2$ molecules on alkali metal surfaces have been explored extensively and are relatively well understood in terms of the least attractive adsorption interactions in nature. Much less thoroughly investigated are wetting transitions of Hg, $H_2O$, heavy inert gases and other molecular films. The basic idea is that nonwetting occurs, for energetic reasons, if the adsorption potential's well-depth D is smaller than, or comparable to, the well-depth $\varepsilon$ of the adsorbate-adsorbate mutual interaction. At the wetting temperature, $T_w$, the transition to wetting occurs, for entropic reasons, when the liquid's surface tension is sufficiently small that the free energy cost in forming a thick film is sufficiently compensated by the fluid-surface interaction energy. Guidelines useful for exploring wetting transitions of other systems are analyzed, in terms of generic criteria involving the "simple model", which yields results in terms of gas-surface interaction parameters and thermodynamic properties of the bulk adsorbate.

I. *Introduction*

The physics of liquid film wetting of solid surfaces has long been of fundamental interest and relevant to numerous technologies. The phenomenon of a wetting transition, *per se,* is of more recent interest. In 1977, independent predictions by Cahn and by Ebner and Saam indicated that (thermodynamic) wetting transitions are expected, quite generally, in the case of very weakly attractive gas-surface interactions V(**r**)[1-4]. Such a transition occurs between a low temperature (T) regime of *nonwetting* behavior and a high T regime of *wetting* behavior. These regimes meet at the *wetting temperature* $T_w$. The distinction pertains to the film coverage present in coexistence with a vapor at saturated vapor pressure (svp), $P_0$ and chemical potential $\mu_{vapor} = \mu_0$ . "Nonwetting" (sometimes called *incomplete wetting*) behavior means that just a thin film is adsorbed at svp, while "wetting" (sometimes called *complete wetting)* refers to the presence of a macroscopically thick film. An equivalent distinction between these situations involves the contact angle θ of a bulk droplet on the surface at svp. In the wetting case, θ =0, meaning that the equilibrium film spreads uniformly across the substrate's surface. In the nonwetting case, a macroscopic droplet will coexist at svp with a thin film, exhibiting the finite angle θ where they make contact.

While the transition between wetting and nonwetting is defined in terms of the properties *at* coexistence, there is a distinct manifestation of this transition at a pressure *below* coexistence, i.e., $P<P_0$. This phenomenon is seen dramatically in adsorption isotherm data of in Fig. 1 for the case of $^4$He on Cs. The three isotherms (with data shown only very close to $P_0$) reveal three very different kinds of behavior. At the lowest T, less than a monolayer is adsorbed for any pressure; this is nonwetting behavior. At the highest T, instead, the coverage increases smoothly with P, diverging as P approaches $P_0$, a case of wetting. At the intermediate value of T~2.1 K, the low P data resemble the nonwetting results at T~1.7 K. However, there occurs an abrupt transition about 0.5% below svp; this is the so-called "prewetting transition", seen at pressure $P_{prewet} \approx 0.995\ P_0$. The coverage jumps essentially discontinuously (by a factor of 8) at this point. For $P>P_{prewet}$, the 2.1 K data are similar to the wetting data at T=3.35 K. This *prewetting* transition receives its name because this striking signature- a coverage jump- of the wetting transition is observed *below* $P_0$. The function $P_{prewet}(T)$ forms a first-order transition line in the fluid's P-T plane, starting at the point $(P_0(T_w),T_w)$ and ending at the prewetting critical point. (In the case of Fig. 1, $T_w$ ~2.0 K). Such a phase diagram has been explored for a number of systems which exhibit the wetting transition. While the details are unique to each case, the global topology of these diagrams is universal. For example, a very different range of conditions is relevant to the case of Hg on sapphire, for which the experimental phase diagram appears in Fig. 2. Compared to the $^4$He case ($T_w/T_c$ ~0.38), when expressed as a ratio, the wetting temperature ($T_w$~1,600 K) of the Hg transition is closer to the bulk critical temperature ($T_c$ ~1,750 K); thus $T_w/T_c$≈0.91 in Fig. 2. Of course, the absolute difference $T_c - T_w$~170 K is much larger for Hg/sapphire than for $^4$He/Cs ($T_c - T_w$ ~3.1 K). A remarkable and common feature of these prewetting transition lines is that their pressures lie within a few per cent of the bulk vapor pressure curve. This fact means that the experimentalist must look quite carefully to see the transition and, therefore, that a good theoretical estimate of $T_w$ is an invaluable guide for this search.

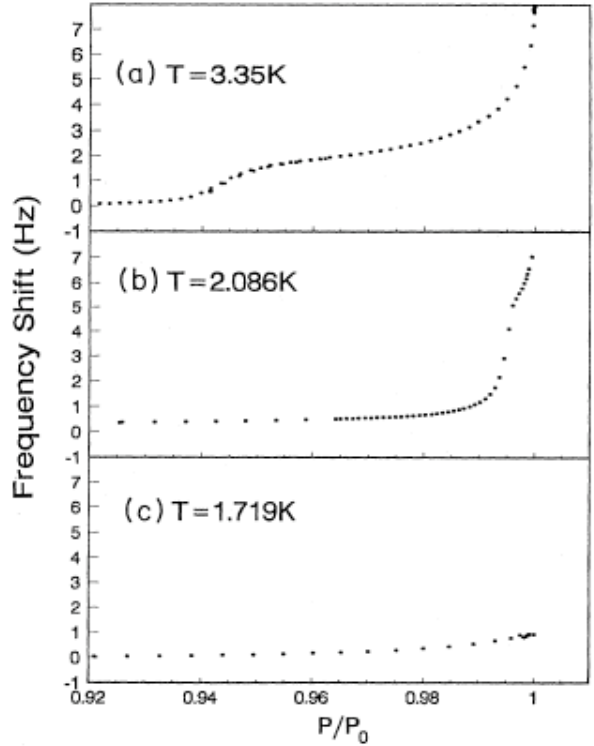

*Fig. 1. Frequency shift (proportional to the film coverage) of $^4$He on Cs, measured with a quartz microbalance, with P close to svp, $P_0$. At low T~1.7 K, the film does not wet the surface, while at high T~3.4 K, it does wet the surface. At intermediate T~2.09 K, the prewetting jump in coverage occurs at $P_{wet}/P_0$=0.995. From Rutledge and Taborek[5].*

Wetting transitions were observed for the first time in the 1990's in these two quite distinct kinds of system: inert gases and $H_2$ on alkali metal surfaces, at low $T_w$, and Hg on various surfaces, with high $T_w$.[5-20] Many other wetting transitions have been predicted, but not yet observed. The general subject of wetting transitions has been reviewed in a number of places, including articles in this volume by Saam, by Ancilotto et al and by Taborek [2, 21-28]. The focus of the present article is the relationship between this transition and the gas-surface interaction, which plays a decisive role in whether and where such transitions occur. At the outset, it should be stated that the low $T_w$ transitions of the inert gases and $H_2$ are *relatively* well understood from first principles calculations, but other transitions are either not understood theoretically or have yet to have the relevant predictions tested experimentally.

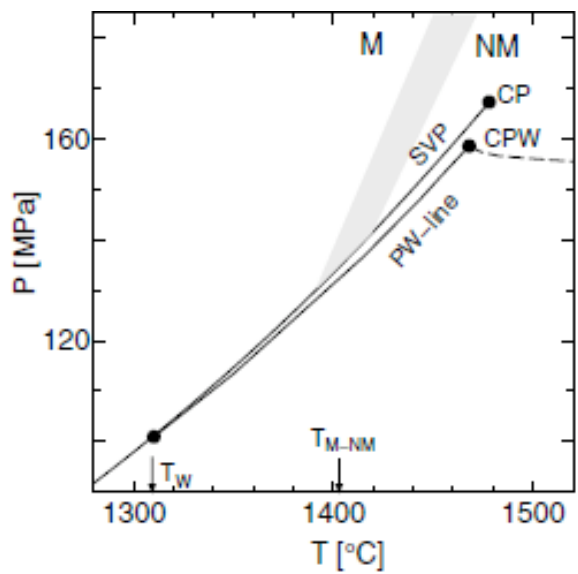

Fig. 2. Prewetting transition line for Hg on sapphire, close to the bulk critical point (CP). Also indicated are the wetting temperature, $T_w$, the prewetting critical point (CPW) and the metal (M) to nonmetal (NM) transition region. From Yao and Ohmasa[6].

Interest in wetting comes from both practical concerns (*e.g.*, adhesion, lubrication, gas storage technologies) and the fundamental desire to understand the *weakest* interactions in nature between atoms and surfaces. As will be discussed in some detail here, wetting transitions are observed when the adsorption energy is smaller than, or comparable to, the cohesive energy within the film. When the film is an inert gas fluid, both of these energies are small. Thus, a quantitative understanding of wetting requires knowledge of these very weak interactions. This need presents a challenge to our understanding of electronic properties at surfaces, which is especially difficult because van der Waals interactions responsible for the attraction involve the *nonlocal* correlation energy- the weak, dynamical coupling between charge fluctuations of the adatom and the solid surface. As such, theories based on local electronic densities and effectively mean-field approximations to the energy functional are usually not sufficient for quantitative purposes[29].

This paper compares behavior for a set of systems for which wetting transitions have been predicted and/or observed. Section II presents a thermodynamic description of wetting, expressed in terms of a surface free energy functional $\sigma(\mu,T,N)$, where the two-dimensional (2D) number density N is to be varied. The minimum of $\sigma$ (as a function of N) occurs at the *equilibrium* film density, at which point the value of $\sigma$ is $\sigma_{sv}$, the solid-vapor interfacial tension. Section III presents a so-called *simple model* of the wetting transition, which provides an implicit, albeit approximate, expression for the wetting temperature if the adsorption potential is known. The model explains why the key feature of systems exhibiting wetting transitions is a weakly attractive adsorption potential, meaning that the adsorption potential's well-depth D is comparable to the well-depth $\varepsilon$ of the mutual interaction between adatoms (or admolecules). Section IV explains the origin of such potentials in qualitative terms, noting that inert gas-alkali interatomic interactions are also quite weakly attractive. Section V presents results concerning wetting for the case of simple gases on alkali and alkaline earth metals, systems which have been studied extensively. Section VI describes wetting transitions for a set of other systems, where the theoretical and/or experimental situation is much less well developed. Section VII briefly summarizes these various wetting problems and suggests directions for future research.

II. *The surface free energy functional $\sigma(\mu,T,N)$*

As shown above, in Fig. 1, one way to observe the wetting transition is an adsorption isotherm. In this experiment, one determines, at fixed T, the coverage N as a function of P, or equivalently the chemical potential $\mu_{vapor}$; for an ideal monatomic gas, this latter quantity is given by

$$\mu_{vapor} = \beta^{-1} \ln (n\lambda^3) \qquad [1]$$

Here, n=P$\beta$ is the 3D vapor number density and $\lambda = (2\pi\beta\hbar^2/m)^{1/2}$ is the de Broglie thermal wavelength, expressed in terms of the atomic mass m and the inverse temperature $\beta=1/(k_B T)$. Such experiments usually cover the range from very low P to $P_0$. One theoretical procedure used to determine the equilibrium value of N is shown schematically in Fig. 3, which presents the "surface tension function", $\sigma(\mu,T,N)$, equal to the grand free energy per unit area, *i.e.,* the difference between the Helmholtz free energy per unit area, $F(T,N)$, and $\mu N$;

$$\sigma(\mu,T,N) \equiv F(T,N) - \mu N \qquad [2]$$

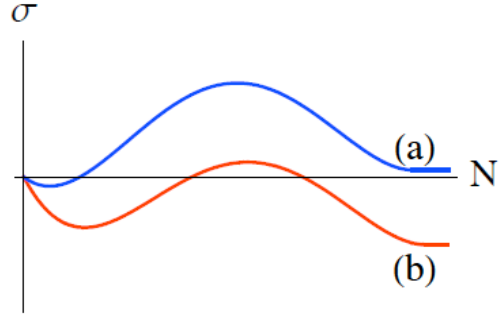

*Fig. 3. Schematic behavior of $\sigma(\mu_0,T,N)$ as a function of coverage N, at svp. In case (a), the global minimum occurs at finite coverage, corresponding to a thin film (nonwetting). In case (b), the minimum occurs at infinite thickness (wetting). Between these two situations there occurs a wetting transition.*

The *equilibrium* state of the film corresponds to the minimum, as a function of N, of $\sigma(\mu,T,N) \equiv \sigma_{sv}$ (at specified T and chemical potential $\mu = \mu_{vapor}$). As stated earlier, the distinction between wetting and nonwetting involves the film's behavior at $P_0$ ($\mu = \mu_0$). Let us suppose that Fig. 3 applies to two different situations, for each of which $\mu = \mu_0$. The behavior at *very* small N is simple to compute (the "Henry's law regime"), because then the film is a quasi-2D ideal gas, for which we may use a relation analogous to that of the vapor, Eq. 1 [28]:

$$(\partial F/\partial N)_{N\to 0} = \mu_{ideal2D}(N,T) = E_1 + \beta^{-1} \ln[N\lambda^2 (1-e^{-\beta\Delta})] \qquad (N\to 0)$$

Here, $E_1$ is the ground state energy of the adatom and $\Delta$ is the spacing between energy levels of perpendicular vibration in the surface potential, $V(z)$, which is implicitly assumed to be a quadratic function of z. Correspondingly, there is an initial slope to the surface tension function, given at svp by

$$[\partial\sigma/\partial N]_{N\to 0} = (E_1-\mu_0) + \beta^{-1} \ln[N\lambda^2 (1-e^{-\beta\Delta})] \qquad (\mu=\mu_0, N\to 0)$$

Neglecting the second term, which is roughly proportional to T, one observes that the initial slope of $\sigma(\mu_0,T,N)$ has a sign equal to that of $(E_1-\mu_0)$ [30]. This difference is assumed to be negative in the figure; meaning that a single atom is bound more strongly to the substrate than to the bulk adsorbate at the given value of T. If, instead, the sign were positive, no such thin film *ever* occurs in equilibrium on this surface, since then $\sigma(\mu_0,T,N)$ always has a local minimum at N=0 for all $\mu_{vapor} \leq \mu_0$. The case of the He isotopes at low T on Cs is quite interesting in this respect. For $^4$He, the atomic binding energy for this potential (discussed in Section IV) is 3.8 K so, with $\mu_0 \approx$ -7.2 K, the difference $(E_1-\mu_0) \approx 3.4$ K. Hence, perhaps surprisingly, the initial slope would be *positive* at low T, meaning *no* monolayer film forms, as seen in the data of Fig. 1. In contrast, the less strongly bound $^3$He atom $(E_1 \approx -3.4$ K) has $(E_1-\mu_0) \approx -0.9$ K, so the initial slope of the function $\sigma(N)$ is negative at low T. Hence, a monolayer film forms at low T only for the *less* strongly bound $^3$He isotope.

The determinant of wetting *vs.* nonwetting is the form of $\sigma(\mu_0,T,N)$ at *large* N. Fig. 2 exhibits two distinct types of behavior for the limiting behavior of this function. Case (a) is nonwetting; only a thin film is present at svp, perhaps a monolayer or so. The alternative scenario -case (b), wetting- occurs when the minimum lies at infinite thickness. (In fact, "infinite thickness" really means a finite coverage, of order 30 to 200 nm, once gravity is taken into account.) The wetting transition occurs between situations (a) and (b). This variation of behavior is usually achieved by changing T, as in Fig. 1, although one could alternatively make a quasi-continuous variation of the substrate composition or structure to achieve the same goal[8,31-34], as discussed briefly in Sect. V. The temperature of this transition is $T_w$, at which point the equilibrium film coverage at svp jumps from a finite value to infinity- a first order transition. A continuous transition can occur for some systems, in principle[35], but that possibility has not yet been seen experimentally in adsorption on solid surfaces. However, it has been seen in adsorption on liquid surfaces[2,36,37].

In the case of a macroscopic wetting film, the surface free energy manifestly coincides with the sum of the solid-liquid ($\sigma_{sl}$) and liquid-vapor ($\sigma_{lv}$) surface tensions (each of which is a function of T). Thus,

$$\sigma_{sv} = \sigma_{lv} + \sigma_{sl} \qquad [3] \qquad \text{(wetting case)}$$

In the nonwetting case, instead, a thin film forms, as seen in Fig. 1c and Fig. 3a. This behavior means that the corresponding solid-vapor tension is smaller than the sum on the right side of Eq. 3: $\sigma_{sv} < \sigma_{lv} + \sigma_{sl}$. The inequality is consistent with Young's equation, which provides the contact angle $\theta$:

$$\sigma_{sv} = \sigma_{lv} \cos\theta + \sigma_{sl} \qquad [4] \qquad \text{(nonwetting case)}$$

Evidently, wetting corresponds to the limit of zero contact angle. The spreading coefficient w is defined as the difference between the two sides of Eq. 3:

$$w \equiv \sigma_{sv} - (\sigma_{lv} + \sigma_{sl})$$

If w<0, a droplet beads up on the surface; otherwise, it spreads across the surface. Physically, the former situation is one for which the solid-vapor interface has the lowest free energy, so the area of solid-vapor contact is maximized as the liquid equilibrates.

III. *The "simple model" of the wetting transition*

While computer simulation is usually the theoretical tool of choice for quantitative calculations of wetting behavior, one might like to have a more convenient method to anticipate whether a given system is wetting, or not, and, if not, at what $T_w$ the transition is expected to occur. We have found useful a so-called "simple model" [8,29,38-41] in which one *approximates* the solid-liquid (s-l) interfacial tension as follows:

$$\sigma_{sl} \approx \sigma_{sv} + \sigma_{lv} + \rho_l \int dz \ V(z) \qquad [5]$$

Here, the adsorption potential is taken to be a function of just the surface-normal distance, z; this is a reasonable approximation for the weakly attractive potentials responsible for nonwetting behavior at low T, since the adsorbed atom lies far from the surface atoms. The right-most term in Eq. 5 approximates the gas-surface interaction energy in terms of the bulk liquid density $\rho_l$ and the integral between the minimum in the adsorption potential, $z=z_{min}$, and $z=\infty$. The physical content of Eq. 5 is that the total free energy cost of the s-l interface ($\sigma_{sl}$) equals that of terminating the solid ($\sigma_{sv}$) plus that of terminating the liquid ($\sigma_{vl}$), with a "correction", the last term, due to the solid-liquid interaction energy. With this approximation, Eq. 5, Eq. 3 leads to an implicit equation for the wetting temperature:

$$[\sigma_{lv}/\rho_l]_{Tw} = -(1/2)\int dz\, V(z) \qquad [6] \qquad \text{(simple model's transition condition)}$$

A very similar analysis and its implication for wetting at T=0 can be developed straightforwardly, *without* any approximation, apart from use of a specific lattice-gas model. In the model's simplest form, the atoms occupy sites of a simple cubic lattice, with nearest-neighbor interactions -ε. Then, the bulk chemical potential equals the cohesive energy, $\mu_0 = -3\varepsilon$, since each particle has six neighbors, and the l-v surface tension is $\sigma_{lv} = \varepsilon/(2a^2)$, where $a$ is the lattice constant (as can be found by calculating the excess energy of a free-standing slab of liquid). The substrate provides an attraction $V_j$ for film atoms at distance $ja$. Then, the criterion for a wetting film to exist at svp is that the grand potential per unit area, given by Eq. 2, be negative; at T=0, then,

$$0 > E - \mu_0 N = \varepsilon + \sum_{j=1}^{\infty} V_j \qquad [7] \qquad \text{(wetting criterion for lattice model at T=0)}$$

This equation coincides with the discretized limit of the continuum simple model, Eq. 6. An even *simpler* hand-waving argument, leading to the same wetting criterion, Eq. 6, at finite T, is this: an infinite film costs free energy $2\sigma_{lv}$ plus the integral in Eq. 5. Wetting occurs only when this sum is negative definite. An equivalent interpretation in thermodynamic terms is that the transition occurs when $\sigma_{lv}$ is sufficiently small that the film spreads in order to gain the gas-surface attractive energy (even if this quantity is small). The simple model can be made *less* simple, but more accurate, close to the bulk critical temperature $T_c$ by replacing $\rho_l$ with $\rho_l - \rho_v$, where $\rho_v$ is the coexisting vapor density.

Eq. 7 is particularly useful because it can yield a *crude* criterion for T=0 wetting in terms of the relevant interactions. If we set $-V_1 = D$, the well-depth for the adsorption potential, and neglect all terms beyond the first term in the sum, the wetting criterion Eq. 7 becomes D/ε>1. A more accurate estimate could be to assume that terms beyond the first yield ~ -0.2 D, resulting in a T=0 wetting criterion D/ε>0.8. Although an *even more* reliable estimate is provided in the following section, one can recognize from the preceding approximations that the criterion distinguishing between wetting and nonwetting has a plausible qualitative aspect: a comparison between the energy of the film's adhesive interaction (D) and its cohesive interaction (ε). The estimate of the wetting temperature, Eq. 6, however, requires additional knowledge- that of the surface entropy of the fluid, which is incorporated in the T-dependence of $\sigma_{lv}$.

The simple model is valuable because of its *prima facie* simplicity. As discussed below, detailed comparison has been made with computer simulations and density functional calculations of wetting transitions[8,21,23,34,39-51]. It is found that the simple model, Eq. 6, provides a rather good approximation to $T_w$, in general, surprisingly so in view of its simplicity.

IV. *Ultraweak adsorption potentials*

What should be borne in mind is the uncertainty of *every* prediction of wetting transitions. The "weakest link" is usually the adsorption potential, which is usually not well known, except at large z. This asymptotic behavior of the attractive potential is a van der Waals interaction, the form of which *is* a known function:

$$V(z) \sim -C_3/z^3 \qquad\qquad [8]$$

The coefficient $C_3$ is calculable in terms of the dielectric function of the substrate, $\varepsilon(\omega)$, and the frequency-dependent polarizability of the adsorbed gas, $\alpha(\omega)$. If this information is available, the resulting value of $C_3$ may be computed, with typical uncertainty 10%, due to uncertain input data. The following approximation is correspondingly accurate for many situations [28]:

$$C_3 = (g_0\alpha_0\hbar\;\omega_s/8)(1+\omega_s/\omega_0)^{-1} \qquad\qquad [9]$$

Here, $g_0$ is the low frequency limit of the surface response function, $g(\omega) = [\varepsilon(\omega)-1]/[\varepsilon(\omega)+1]$ , which is adequately described by this expression (at imaginary frequency i$\omega$): $g(i\omega) = g_0/(1 + \omega^2/\omega_s^2)$. An analogous expression is frequently used to describe the polarizability: $\alpha(i\omega) = \alpha_0/(1 + \omega^2/\omega_0^2)$, where $\alpha_0$ and $\omega_0$ are the static value and characteristic frequency, respectively, of the adatom's polarizability. Eq. 9 provides a useful estimate if the necessary input information about $\varepsilon(i\omega)$ and $\alpha(i\omega)$ is incomplete. Tabulated values of $C_3$ and the parameters $\omega_0$, $\alpha_0$, $g_0$ and $\omega_s$ for many systems appear in Bruch et al, Cheng et al and Vidali et al.[28,39,52]

The real challenge in computing these potentials is the repulsive part of the interaction, arising from charge overlap of the electronic wave functions of the substrate and adsorbate. A set of detailed *ab initio* calculations of V(z) has been carried out by Chizmeshya et al, abbreviated CCZ [38]. Their method combines a Hartree-Fock repulsion, based on a pseudopotential scattering theory of Chizmeshya and Zaremba,[53,54] with a first principles evaluation of the damped van der Waals attraction, where the term "damped" refers to a small z correction to the otherwise divergent Eq. 8. Fig. 4 shows the CCZ potential for a He atom interacting with the surface of Cs. The well-depth of this adsorption potential is just D=0.59 meV, or 6.9 K. This value is significantly smaller than that (ε≈11 K) of the He-He interatomic potential, which is also shown in Fig. 4.[55] It is intriguing that the most (chemically) reactive surface is the most inert surface for physisorption!

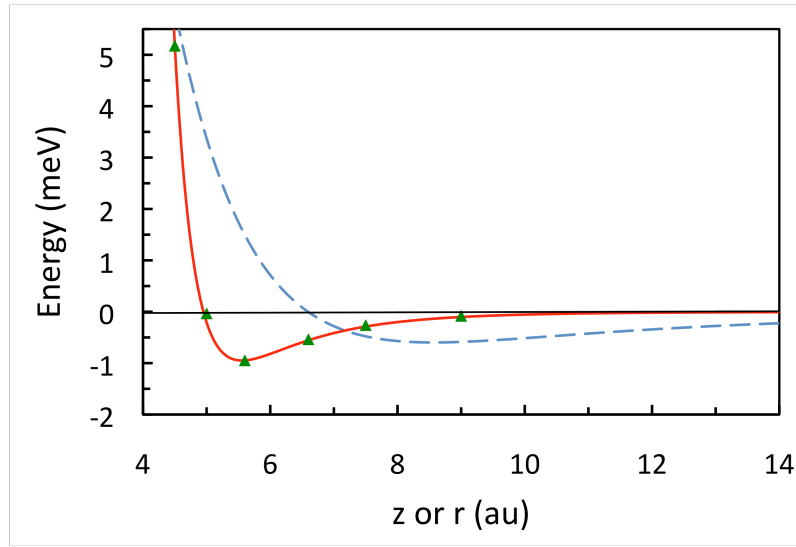

*Fig. 4. Comparison between the He-He interatomic interaction (full curve) and the adsorption potential (dashed curve) for a He atom on Cs metal, where 1 meV≈11.6 K. The distance z (expressed in Bohr radii=au≈0.529 Å) is measured with respect to the jellium edge, which is approximately one-half of a lattice constant outward from the first layer of Cs nuclei.[38] Individual points (triangles) represent the He-He interaction, from Anderson[55], while the full curve is a Lennard-Jones potential, with parameters ε= 0.949 meV (11 K) and σ= 4.95 au.*

In view of data in Table I and the *estimated* wetting criterion discussed above, D/ε>0.8 , it is not surprising that $^4$He does not wet Cs at T=0. The real situation is more complicated, as we shall see; $^3$He films *do* wet Cs at all T[15]. Although this may seem paradoxical, because a single atom of $^4$He is bound *more* strongly than a $^3$He atom, $|E_1^{(4)}|>|E_1^{(3)}|$, a key quantity (as mentioned in Sect. II) is the difference  between these energies and the bulk chemical potentials, *i.e.* $(E_1-\mu_0)$, which is negative only for $^3$He.

Table 1 presents values of the ratio D/ε for various adsorption systems, all of which are expected to exhibit nonwetting behavior at the triple point, except those involving graphite and probably Kr, Xe and $H_2$ on Mg[56]. Apart from these exceptions, the ratios are small. A general feature of the data is particularly low ratios for gases adsorbed on Cs and Rb, followed by progressively higher values for the heavier alkali metals and Mg, the one alkaline earth metal in the table. The ratios are much higher for graphite, which is the *most* attractive substrate for many physisorbing gases. Note that the ratios are even smaller for Ne than for He, meaning that Ne is particularly likely not to wet most of these surfaces, as confirmed below in a more reliable analysis based on the simple model, or computer simulations, or experiments.[12,34,41,50]

TABLE I

*The ratio D/ε, for various gas-surface combinations. ε values are from Bruch et al [28] and D values are from CCZ[38] or Vidali et al [52], for graphite. For each metal its work function W, in eV, is in parentheses; values taken from Refs.[57] and [58].*

Gas (ε, in K, in parentheses)

|  | He | Ne | $H_2$ | Ar | Kr | Xe |
|---|---|---|---|---|---|---|
| Substrate (W) | (11.01) | (42.25) | (34) | (143.2) | (201.3) | (282.8) |
| Cs (2.14) | 0.63 | 0.56 | 0.70 | 0.94 | 1.1 | 1.1 |
| Rb (2.26) | 0.66 | 0.57 | 0.71 | 0.93 | 1.1 | 1.1 |
| K (2.30) | 0.74 | 0.61 | 1.3 | 0.99 | 1.1 | 1.1 |
| Na (2.33) | 1.1 | 0.88 | 2.0 | 1.3 | 1.5 | 1.5 |
| Li (2.93) | 1.6 | 1.2 | 2.9 | 1.7 | 1.9 | 2.0 |
| Mg (3.66) | 3.2 | 2.2 | 5.6 | 2.9 | 3.2 | 4.0 |
| Graphite (5.0) | 18 | 9.0 | 18 | 7.8 | 7.2 | 6.6 |

The origin of the weakest adsorption interactions in Table 1 is that the conduction electrons of the alkali metals have wave functions extending far outside these metals' surfaces, repelling the inert gas atom. There is a sensible correlation with the work function W; alkali metals have the smallest work functions in the periodic table. Hence, their electronic charge density $\rho_e(z)$ decays most slowly outside of the surface. Asymptotically, $\rho_e(z) \propto \exp[-Kz]$, with $K=(8m_eW)^{1/2}/\hbar$ and $m_e$ is the electron mass; for example, for Cs (W≈2.1 eV), the decay constant K≈1.5/Å. This slow decay and correspondingly extended repulsive interaction [38] leads to a large equilibrium distance, of order 6-8 Å, measured from the outermost layer of nuclei, for the adatom and therefore weakly attractive potentials, according to Eq. 8. In analyzing the weak attraction, one should also understand the role played by the coefficient $C_3$. Its value is 1.4 (1.9) times larger for He/Li (Mg) than for He/Cs. These ratios are not, by themselves, sufficient to account for the 3.3 (5.5) times larger values of D for He on Li (Mg) than on Cs. The reason is a positive feedback in the potential; the greater $C_3$ values for the heavier alkali metals reduce the equilibrium distance, further increasing D. We note, based on these arguments, that other alkaline earth metals should also manifest weak attractions for inert gases; Ba, for example, has W=2.6 eV, implying that it should be much less attractive than Mg. However, no such adsorption potential calculations or experiments have been carried out with this surface, as far as we know.

Fig. 5 presents an alternative way to think about these adsorption potentials. Consider the adsorption potential near a surface for which the decay length of the surface charge density, L=1/K. Then, if the repulsion is proportional to this density,[59,60] L provides the characteristic length scale for the full potential V(z) (since the van der Waals interaction has no range *per se*). Hence, L should be proportional to the characteristic length for the full potential, $z_0=(C_3/D)^{1/3}$. Fig. 5 reveals such a correlation, approximately, for a number of gas-surface interactions, in that the data for quite different systems are, indeed, obeying roughly a linear relationship.

The following section describes how wetting behavior serves as the principal tool to assess theoretical adsorption potentials for these very weakly attractive cases. The general uncertainty in the adsorption potential could be remedied, in principle, if one could use atomic/molecular beam scattering to observe bound-state resonances (selective adsorption)[52]. In this method, the source of the best-known adsorption potentials for other systems, the incident particle is diffracted into a resonant bound state on the surface, dramatically reducing the specularly reflected intensity whenever such a resonance occurs. A kinematic analysis of these data directly yields the spectrum of vibrational eigenstates in the potential V(z). Unfortunately, the small corrugation (lateral periodic variation of the adsorption potential) on alkali metals means that atomic diffraction is extraordinarily small; to date, as a result, no scattering experiment has yielded

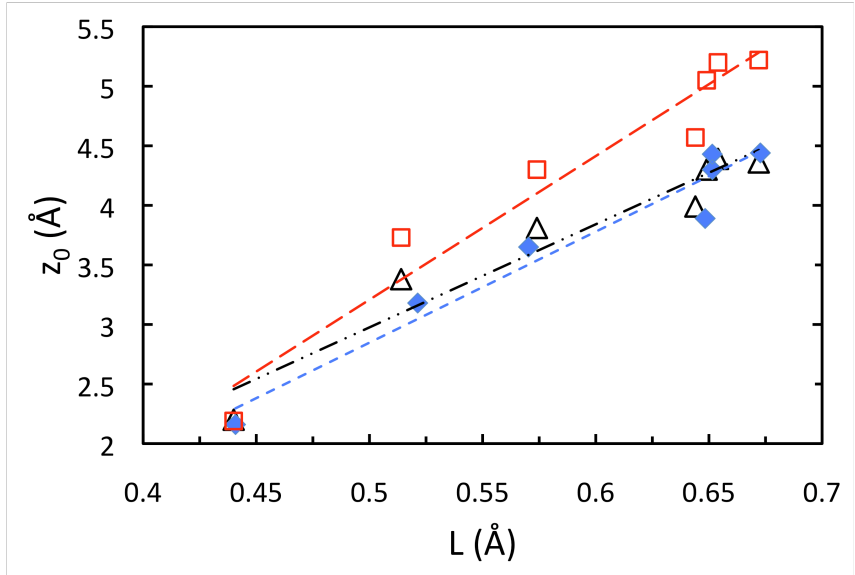

*Fig. 5. Correlation between the characteristic length scale for the adsorption potential, $z_0$, and the decay length for the electronic charge density, L, both defined in the text. Points for various surfaces are derived from data of Table 1 for the gases He (red squares), $H_2$ (blue diamonds) and Ne (black triangles). Lines are linear fit to the data: blue short dashes ($H_2$), red long dashes (He) and black dotted-dashed (Ne)*

bound state resonance information for these weak-binding systems. A less direct technique, rotationally mediated selective adsorption[61,62], *is* feasible; the one experiment thus far [63] found consistency with the conclusion of Section 2, based on simulations, that the $H_2$/Cs interaction is somewhat more attractive than the CCZ prediction of that potential.

Precisely the same physics underlies the set of inert gas-alkali interatomic potentials. For A-He interactions, where A is some atomic species, this can result in strikingly small well-depths, $\varepsilon$, of order 10% of $\varepsilon_{He-He}$, as seen in Table II for several such interactions. There is also seen an expected correlation between the value of $\varepsilon$ and that of the ionization energy $I_A$, analogous to the dependence of D on W for the adsorption potential seen in Table I. These well-depths are quite

TABLE II

*Parameters of A-He interatomic interactions, where A is a listed atomic species: ε is the well-depth and $R_{min}$ is the equilibrium distance for this dimer. Listed in order of increasing $I_A$, the ionization energy of atom A. Potential energy data from Refs.[64-68] for other gases than H-He, which is taken from Bhattacharya and Anderson[69]*

| Atom A | $I_A$ (eV) | $R_{min}$(Å) | ε (K) |
|--------|-----------|--------------|-------|
| Cs | 3.9 | 7.9 | 0.2-2.2 |
| Rb | 4.2 | 7.5 | 0.3-2.2 |
| Li | 5.4 | 6.2 | 1.5-4 |
| Mg | 7.6 | 3.9 | 6.5-11 |
| H | 13.6 | 3.7 | 6.8 |
| He | 24.6 | 2.97 | 11.0 |

uncertain, however, due to open questions about the damping of the van der Waals attraction ($\sim r^{-6}$ asymptotically). In the case of He-He, in contrast, essentially exact calculations are available, so the resulting well-depth is known to better than 1%.[55]

V. *Wetting transitions on alkali and alkaline earth metals*

The first wetting transition to be observed experimentally was that of $^4$He/Cs, which had been predicted with both the simple model, Eq. 6, and a more detailed helium density functional theory[5,7,8,16,17]. These experiments included measurements of superflow and third sound, which probe the film thickness indirectly through the superfluid properties, as well as surface plasmon microscopy, ellipsometry and quartz microbalance measurements, which provide more direct evidence of the film coverage. Subsequent experiments have found similar wetting transition behavior of $^3$He, $H_2$, $D_2$ and Ne on a variety of alkali metal surfaces. These experiments, as well as those of the contact angle, on alkali metals are described in articles of Taborek and Saam in this volume.[22,23]

One can investigate these transitions with approximate models, like those discussed in Sect.III, or with more reliable computer simulations. An example of the latter appears in Fig. 6. Here, one observes nonwetting behavior of $H_2$ on Rb at 25K; the surface-excess film coverage is seen to be small, about 15% of a monolayer or less, for all P< $P_0$. In contrast, at 26 K, there occurs a discontinuous jump in coverage (by a factor ~8) to a multilayer film, at P*= $P_{prewet}$(26K)/$P_0$ ≈0.97. Beyond this value of P*, the coverage grows smoothly with increasing P*, *i.e.,* wetting behavior. This means that the (predicted) wetting temperature is $T_w$ =25.5 ±0.5 K. However, the experimental value found by two groups [9,10,37] is $T_w$ ≈19 K. Shi et al suggested that the discrepancy in the value of $T_w$ may arise from a well-depth for $H_2$/Rb that needs to be increased by some 20% above the *ab initio* value computed [38] by CCZ.

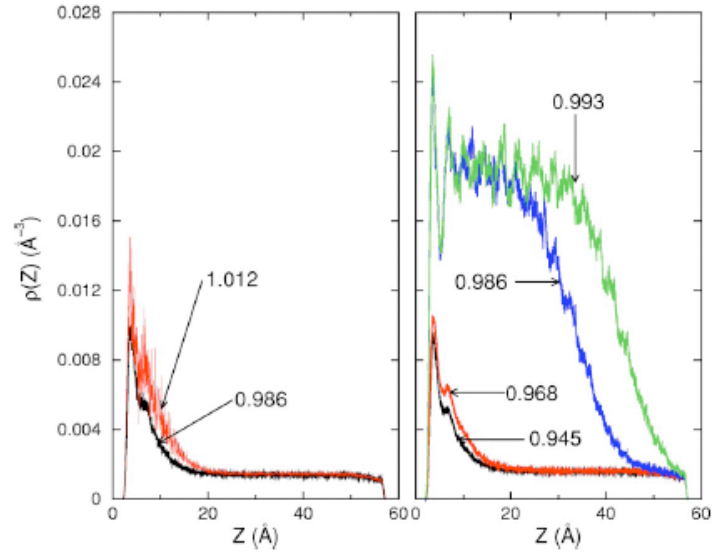

*Fig. 6. Number density ρ(z) of $H_2$ molecules as a function of distance z from the Rb surface (measured from the surface layer of nuclei). Left panel is at simulation temperature T=25 K, showing nonwetting, while right panel is at 26 K, showing the prewetting transition. Numbers within the figures represent the saturation ratio, P*=P/$P_0$. The density falls to zero at the right end of the simulation domain, at z=60 Å. From Shi et al[33].*

Much easier to carry out are investigations based on the simple model. Fig. 7 presents such a comparison (adapted from Kim et al [70]) between interaction parameters for $^4$He and various surfaces and the criteria for monolayer formation and wetting, respectively, at T=0. The wetting criterion is based on Eq. 6. For this purpose, it is convenient to evaluate that relation with an approximate "3-9" adsorption potential:

$$V(z)/D = (4/27)x^{-9} - x^{-3} \qquad [10]$$

Here, x= z/z₀, where $z_0=(C_3/D)^{1/3}$ is the characteristic length scale in the potential. This potential has its minimum at $z_{min}/z_0=(2/3)^{1/3}\sim0.87$; its full-width at half minimum is 0.485 $z_0$. The wetting criterion Eq. 6 for this potential then becomes[8]

$$(C_3D^2)^{1/3}= 3.33 \ (\sigma_{lv}/n_l) \qquad [11]$$

This equation can be further simplified if one assumes that the interaction, Eq. 10, can be represented as an integral of Lennard-Jones pair interactions (with hard-core diameter parameter $\sigma_{hard}$) over a half-space[70]. In this case, the wetting criterion, Eq. 11, becomes $D\sigma_{hard}$ =41.6 K-Å for ⁴He. Here, $\sigma_{hard}/z_0= 10^{1/6}/3^{1/3} \sim 1.02$.

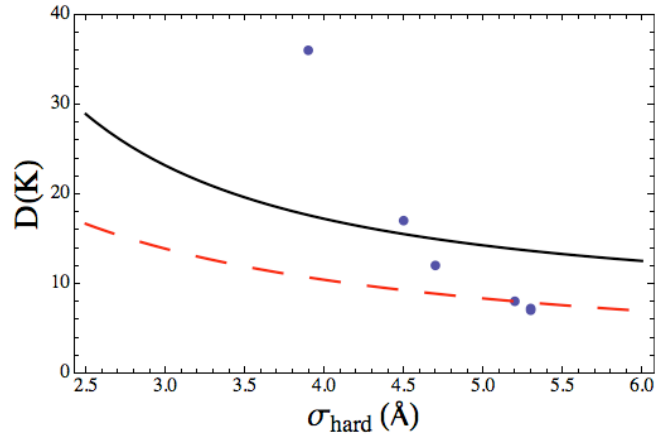

Fig. 7. Threshold values of D for ⁴He monolayer formation (full curve) and T=0 wetting (dashes) on various surfaces, as a function of the hard-core diameter $\sigma_{hard}$ of the He-substrate interatomic potential. Points are derived from CCZ calculations[38] of $C_3$ and D. The substrates are (in order of decreasing D) Mg, Li, Na, K, Rb and Cs.

The monolayer formation criterion appearing in Fig. 7 is that the ground state energy per atom of three-dimensional (3D) helium equals that of a quasi-2D film:

$$E_{3D} = E_{film} = E_1 + E_{2D} \qquad \text{(monolayer criterion)} \qquad [12]$$

Here, $E_1$ is the ground state energy of a single adsorbed atom and $E_{3D}$=7.17 K ($E_{2D}$=0.90 K) is the energy per atom of 3D (2D) [4]He at T=0.[71-73] The energy $E_1$ is computed from the adsorption potentials for various surfaces, taking into account the zero-point energy of motion perpendicular to the surface. In the harmonic approximation to the potential of Eq. 10, the criterion is

$$E_{3D} - E_{2D} = -D + [\hbar/(2\sigma_{hard})][\alpha D/m]^{1/2} \qquad [13]$$

Here, $\alpha=27(5/2)^{1/3}\sim36.6$ and m is the adatom's mass. This expression provides the basis for the monolayer curve in Fig. 7, with $E_{3D} - E_{2D}$ = -6.32 K for [4]He at T=0.

If a given system's value of D falls below a specific line in Fig. 7, the relevant phase is predicted *not* to occur. Direct implications of this figure are then that (at T=0) [4]He does not wet Cs (since the circle lies below the dashed curve), while [4]He/Rb and [4]He /K wetting are marginal (the square and diamond lie essentially *on* the curve). The first prediction is consistent with numerous experiments [5,7,16,17,26] on Cs. As for [4]He/Rb, the borderline theoretical prediction is thus far consistent with experiments, insofar as the data from different groups disagree concerning T=0 wetting [11, 19, 20, 23, 25,74]. The prediction for K is also marginally consistent with experiments showing wetting at T=0 and the absence of a monolayer. Na, in contrast, is predicted by the figure to be wet by [4]He, although it does *not* form a monolayer film, since its symbol ($\Delta$) falls below the full curve [29]. This situation implies the existence of a wetting transition at low T, *i.e.* a line of coverage discontinuity in the $\mu$-T plane. There is, indeed, experimental evidence of this transition on Na, below 1K [19,25]. A similar transition is seen experimentally for K [25]. Finally, the figure indicates that Li is expected [43] to show continuous adsorption at low T. This prediction, too, is consistent with experiments. [68] Thus, the group of five wetting behaviors of [4]He on alkali metal surfaces is consistent with the current wetting theory and theoretical potentials. Making this statement more comprehensive requires use of more reliable and extensive methods than that used to generate the figure above. At T=0, wetting transition calculations for He have been carried out with density functional methods, hypernetted chain and diffusion Monte Carlo methods, of which the last is "exact", in principle (meaning subject only to statistical uncertainty and incomplete information about the interactions)[75]. At nonzero T, limited path integral Monte Carlo calculations exist thus far only for [4]He/Li and Cs [43,70], while the density functional method has recently been extended to nonzero temperatures and applied to

[4]He/Cs [71,72]. The present article is not the place to review these many calculations. We note also that the film coverage greatly influences its superfluid properties, a dependence which has been explored experimentally by the UC Irvine group[22,25,74], but not theoretically to any significant extent.

The preceding section described the origin of He wetting transition in terms of the small value of D and showed that this quantity is correlated with work function. Demolder et al pointed out [76] that oxidation of the Rb surface would lower the work function and hence decrease the well-depth, raising the wetting temperature. Their subsequent experiments confirmed this expectation for the oxidized Rb surface[77]. Experiments found the nonwetting behavior of [4]He to persist up to 1.45 K, much higher than the value ($T_w \sim 0$) for the bare Rb surface.

We stated earlier that [3]He wets Cs at T=0 (exhibiting a prewetting jump), while [4]He does not wet that surface. The origin of this intriguing difference can be seen by comparing the left side of Eq. 6 for the two isotopes. For [4]He at T=0, the left hand-side is $\sigma_{lv}/\rho_l$=13 K-Å, while for [3]He this ratio is just 6.9 K-Å. This means that the wetting well-depth threshold is much lower for the lighter isotope, so that even the modest depth for the He-Cs potential, Fig. 4, easily exceeds the wetting threshold of [3]He. Since we believe that the Cs surface provides He with the least attractive adsorption potential of any surface, we conclude that [3]He is a *universal wetting agent*. It is possible that [3]He is the *only* such adsorbate in nature, but we lack enough information about other systems' adsorption potentials to test this conjecture.

Since it was established that [4]He does not wet the surface of Cs below 2 K, while [3]He does wet that surface, one can ask about the properties of [4]He-[3]He mixtures on Cs. Indeed, Pettersen and Saam predicted that this system would exhibit a reentrant wetting transition, depending on the concentration $x_3$ of the mixture.[78,79] Experiments in several groups have confirmed this general picture, resulting in a rich phase diagram in the T-$x_3$ thermodynamic plane [80-83]. Among these results is the presence of a dewetting transition on the low $x_3$ side of the coexistence curve, in addition to a prewetting transition on the high $x_3$ side of this curve.

We now turn to the problem of $H_2$ wetting. Thus far, detailed simulations have been performed only for $H_2$ films on Rb and Cs (as shown in Fig. 5) surfaces, as well as thin films of these metals adsorbed on Au surfaces. To develop a quantitative, albeit approximate, overview we employ the same model used to construct Fig. 7 for the wetting behavior of [4]He. Fig.8 presents results for the wetting criterion and monolayer formation criterion at the triple point (13.8 K) of $H_2$. The wetting line corresponds to the relation $D \sigma_{hard} = 3.39 (\sigma_{lv}/n_{triple}) = 283$ K-Å; this coefficient is much larger than that (41.6 K-Å) for $^4$He, primarily because of the much lower $\sigma_{lv}$ of $^4$He (which is smaller than that of $H_2$ at its triple point by a factor of nearly 8, while the density is just 10% lower than that of liquid $H_2$). The other curve shown in Fig. 8, the monolayer formation criterion, is based on Eq. 13, with $E_{3D} - E_{2D} = -66.4$ K.[71] Implicitly, we are neglecting temperature-dependent contributions to the free energy. According to this figure, $H_2$ should not form a monolayer on any of the alkali metals, but it should form a monolayer on Mg. $H_2$ is predicted to exhibit a wetting transition at the triple point on Li, but not on the heavier alkali metal surfaces. The experimental data on Rb and Cs are consistent with these predictions, as mentioned above, while predictions for the other systems have not yet been tested experimentally.

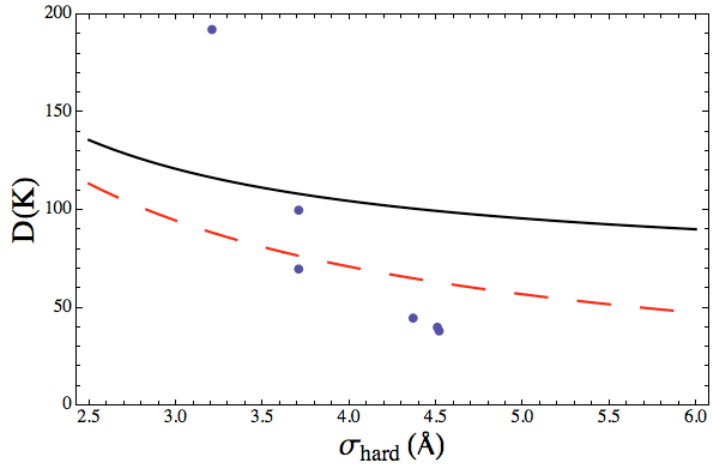

*Fig. 8. Wetting threshold (dashes) and monolayer formation (full curve) thresholds for $H_2$ at its triple point. Data points for various substrates derived from V(z) calculations of CCZ.[38] The substrates are (in order of decreasing D) Mg, Li, Na, K, Rb and Cs. The parameters for Cs, Rb and K nearly coincide.*

There exists the possibility of "tuning" the wetting transition, by creating a "compound" substrate. Consider, for example, a surface consisting of a Rb film of thickness d deposited on Au. Then, if d is large enough that the effect of the Au on the $H_2$ potential is a long-range van der Waals interaction, one can describe the $H_2$ potential energy $V_{Rb/Au}(z)$ as follows:

$$V_{Rb/Au}(z) = V_{Rb}(z) - \Delta C_3/(z+d)^3 \qquad [14]$$

Here, $V_{Rb}(z)$ is the potential energy for $H_2$ molecule at position z above the surface of semi-infinite Rb metal, while the last term represents the "perturbation" attributable to the substitution of Au for Rb at distances from the molecule exceeding (z+d). The quantity $\Delta C_3 \equiv C_3(H_2/Au)$ - $C_3(H_2/Rb)$, *i.e.,* the difference between $C_3$ coefficients for these surfaces. The effect of this perturbation on wetting behavior can be surprisingly large. For example, the $H_2$ potential due to a 15 Å Rb film on semi-infinite Au is barely different from that due to semi-infinite Rb, but the calculated wetting temperature shift is $\Delta T_w = 1$ K, compared to the original $T_w = 19$ K, according to both the simulations and a simple perturbation theory based on the simple model [8]. In the latter approach, the shift on the left side of Eq. 6, proportional to $\Delta T_w$, is equated to the shift of the right side due to the Au perturbation[36]. Similar predictions have been made for the $^4$He/Cs system, which were confirmed experimentally by Taborek and Rutledge[32]. In that case, a Cs film with d~15 Å had $T_w$ ~1.5 K, while infinite Cs has $T_w \sim 2$K. The wetting temperature of $^4$He is particularly sensitive to even small perturbing potentials, according to the simple model, because its value is determined primarily by the T-dependence of $\sigma_{lv}$, which is small for $^4$He at low T (since $\sigma_{lv} \propto T^{7/3}$).[84]

A further variant of the problem of the $H_2$/Cs wetting transition is the effect of adding He in small concentrations to the $H_2$ liquid. This problem has been explored experimentally and theoretically by Pettersen et al.[85] These results are mutually compatible if one incorporates thermal expansion of the fluid and the equation of state of the He impurity within the $H_2$ liquid.

Finally, we turn to the case of Ne adsorption on alkali and alkaline earth metal substrates. Using the same model applied above to He and $H_2$, the wetting criterion, Eq. 11, becomes $D \sigma_{hard} = 367$ K-Å, as shown in Fig. 9. The monolayer formation criterion is again Eq. 13, with $E_{3D}$ - $E_{2D}$ = -151 K [71]. The data points indicate that Ne should adsorb less than a monolayer film at its triple point and wet none of the six substrates represented in the figure. These conclusions (consistent with expectations based on interaction ratio data in Table I) are borne out in the case of Rb and Cs by the experimental data of Hess, Sabatini and Chan.[12] Their data for Rb appears in Fig.10, showing a wetting transition about 0.1 K below the bulk $T_c = 44.4$ K.

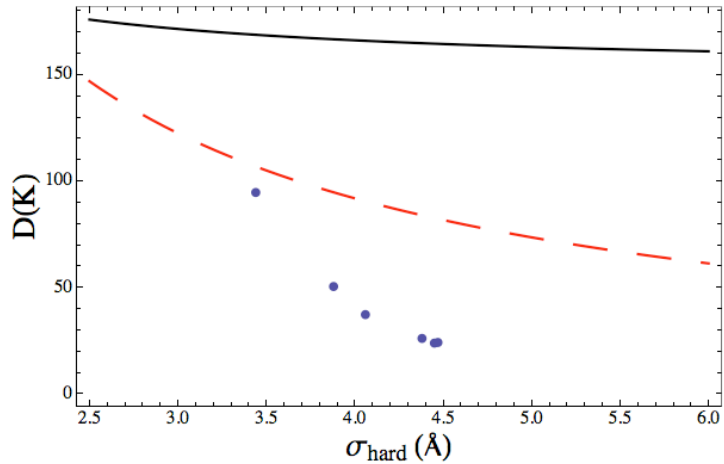

*Fig. 9. Wetting threshold (dashes) and monolayer formation (full curve) threshold for Ne at its triple point, 24.55 K. Data points for various substrates derived from potentials of CCZ. In order of decreasing D, points correspond to Ne/Mg, Li, Na, K, Rb and Cs, respectively.*

Calculations have been carried out for Ne adsorption on alkali and alkaline earth metals, using both Monte Carlo simulations and density functional methods[45,48,50]. The calculations for Li, Mg, Rb and Cs yield results consistent with Fig. 9 and reasonably close to the experimental data

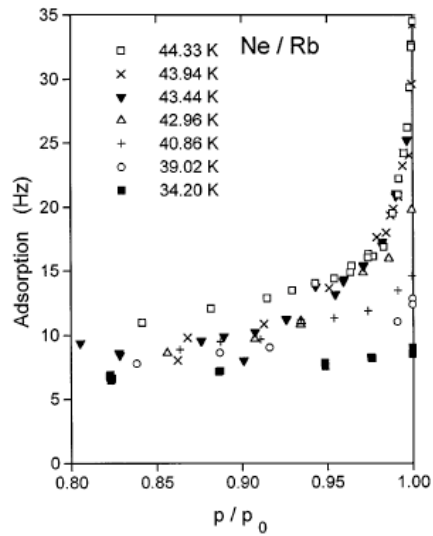

*Fig. 10. Wetting behavior of Ne on Rb close to its critical temperature (~44.4 K), with coverage measured as resonant frequency shift of the quartz microbalance. The wetting temperature is determined to be $T_w$=44.3 K. Figure from Hess, Santini and Chan.[12]*

of Fig. 10 for Rb. These simulation results are also consistent with the experimentally observed "drying behavior" for Cs[12,48]. Agreement with the experiments can be improved if the adsorption potential's well-depth D is reduced by about 9% from the value found by CCZ. This shift is of the *opposite* sign to that needed to explain the $H_2$/Rb wetting transition data. The "drying behavior" of Ne on Cs, is not a drying transition *per se*, but is instead a nonsingular reduction in adsorbed mass close to the surface, within a distance that grows with the correlation length, divergent at $T_c$. This generic behavior was anticipated by Ebner and Saam,[86] who found that long-range interactions move what would be a drying transition (for short-range interactions) to the critical point.

Extensive Monte Carlo calculations were also carried out for Ne adsorption at the surface of Li and Mg. In the former case, the simulation data established that no wetting transition occurs for T<42 K, consistent with Fig. 7, but we may tentatively extrapolate from the Rb data to predict a transition for Ne/Li between 42 and 43.5 K. As for Mg, a wetting transition was found near $T_w$ =28 K. This latter prediction has not yet been tested experimentally.

VI. *Wetting transitions on other surfaces*

In this section, we address briefly the nature of wetting transitions on surfaces other than alkali and alkaline earth metals. We mention four quite different examples to demonstrate the generality of the phenomenon. While all but one of these has been explored experimentally, to some extent, none is understood theoretically. In addition to these four, *many* other systems are predicted to show similar transitions, so there remains much experimental and theoretical work to be done.

The first of these to be discussed is actually the first system proposed by Ebner and Saam [4] a likely candidate to exhibit a wetting transition: Ar on $CO_2$. This case was subsequently found, in experiments of Mistura et al, *not* to manifest this transition, but instead to exhibit triple point wetting[87]. The situation was explained[88] in terms of a more realistic interaction model than was originally used by Ebner and Saam[4]. However, a closely related system, Ne/$CO_2$, was predicted by the same theoretical approach to *have* such a wetting transition, some 5K below the critical point of Ne[80]. Unfortunately, subsequent experiments of Bruschi et al did *not* confirm this prediction, instead showing triple point wetting of Ne on $CO_2$.[89] Thus, it must be said that we do not adequately understand this class of adsorption system.

The second system to be discussed was mentioned in the introduction; wetting transitions have been seen in the adsorption of Hg remarkably close to svp, near its bulk critical point ($T_c \approx 1,750$ K, $P_c \approx 170$ MPa). As evidenced by this temperature scale, these are completely different systems from the inert gases on alkali metals! Fig. 2 shows the phase diagram for this transition on sapphire. While the Hg-Hg interaction is relatively well understood at more moderate temperatures[90], the problem becomes quite complex in this region of the phase diagram because the metal-nonmetal transition occurs quite close to the critical point, so that use of a simple interatomic potential model is questionable there. The data in Fig. 2 were obtained from anomalies in the optical reflectivity. Similar wetting transition behavior has been observed for Hg on Mo and Nb, using acoustical methods, by Kozhevnikov et al.[14,18] Intriguingly, perhaps, the prewetting critical point for Hg/Mo occurs at a *higher* temperature than the bulk critical temperature. This possibility may be a surprise for those experienced with nearest-neighbor Ising models, for which 2D critical temperatures are lower then 3D values because of the smaller coordination number in 2D. However, there is no fundamental reason why a similar difference ($T_{c,3D} > T_{c,2D}$) between these temperatures should be expected for the wetting transition[91].

The third system we mention is water/graphite. The contact angle at room temperature for this system is large, measured variously to be somewhere between 42 and 86 degrees[92-94]. This ambiguity renders uncertain tests of the gas-surface interaction; furthermore, theoretical calculations of the potential yield quite varied results[95-97]. The prediction of a water/graphite wetting transition has been made in two papers, but no complete test of this prediction has yet been made[98,99]. Fig. 11 shows simulation results of Zhao, below and above the transition temperature, which he estimates to be $T_w = 475$ to 480 K. He also identified the prewetting critical point, for which the critical temperature is determined to be 505 to 510 K. These predictions are quite sensitive to the adsorption potential, leaving a question about the accuracy of the predictions. This deficiency provides additional motivation for exploring this system in the future, especially because it is so relevant to many systems of biological importance[100]. Thus far, experiments of Garcia[101] have found *nonwetting* behavior for water/graphite up to 420 K, consistent with the predictions of Zhao et al.

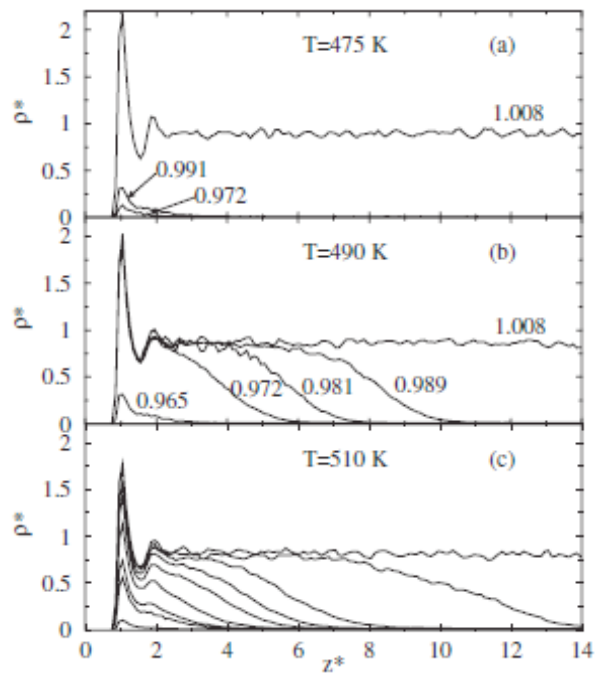

*Fig. 11. Reduced density of water adsorbed on graphite at indicated simulation temperatures, as a function of reduced normal distance, z\*, from Zhao. The wetting transition is thus expected to occur between 475 K and 480 K[99].*

One can construct a "wetting parameter" phase diagram for water analogous to Figs. 6 to 8, for He, Ne and $H_2$. Figure 12 presents this diagram, with one difference from the other figures. The two curves in Fig. 12 for water correspond to wetting thresholds at the triple point and the boiling point, 100 °C, respectively. (These curves are based on the equations $D\sigma_{hard}$ =5,350 K-Å and 4,510 K-Å, respectively.) As can be seen, the water/graphite interaction ($D$ $\sigma_{hard}$ ≈2,610 K-Å) is much too weakly attractive to result in wetting at either temperature, a finding that is consistent with the simulation results of Zhao.

The fourth system we describe in this Section is Xe adsorption on the "compound surface", Cs/graphite, which consists of a monolayer Cs film adsorbed on the surface of graphite. We note that Xe is expected to not wet the surface of *bulk* Cs, at least up to 286 K, close to the critical temperature 289.7 K, according to simulation results of Curtarolo et al[102]; the simulations cannot come closer to $T_c$ because divergent critical fluctuations are not adequately captured in the simulations, due to use of periodic boundary conditions. However, the Cs/graphite substrate provides somewhat greater attraction for Xe than that of semi-infinite Cs, leading to a wetting transition. Figure 13 presents simulation data for this system at 194 K, showing the characteristic prewetting transition behavior at reduced pressure P*~0.97. This prediction of $T_w$, which is not far from the Xe triple point temperature (161.4 K), has not yet been tested experimentally.

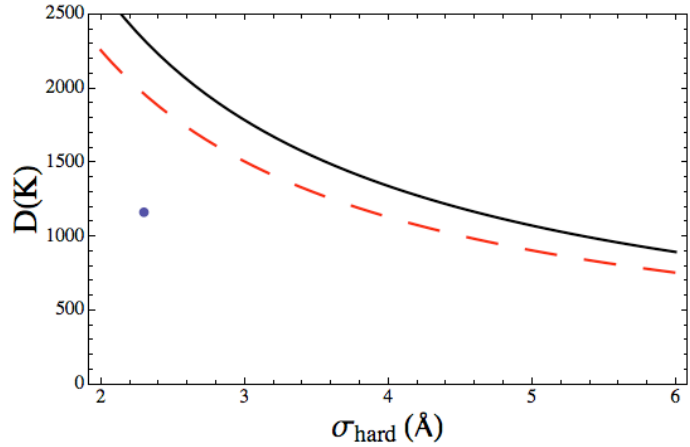

*Fig. 12. Diagram showing wetting criterion for water on various surfaces. Full curve (dashed curve) presents the threshold well-depth at the triple point (boiling point). Indicated point represents the interaction parameters used by Zhao in the water/graphite simulations of Fig. 11.*

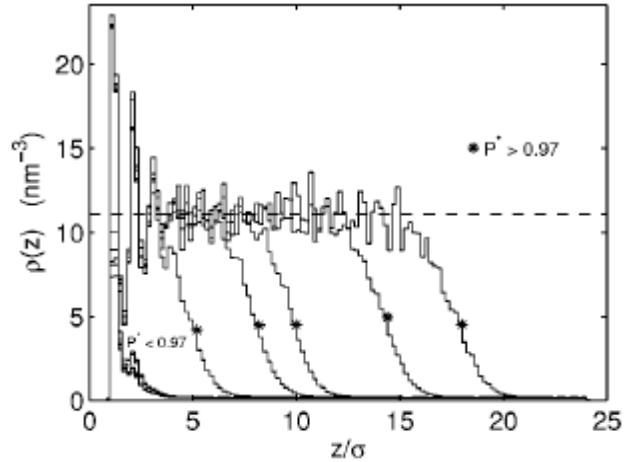

*Fig. 13. Density of Xe on Cs-plated graphite at 194 K as a function of distance, reduced by the hard-core interaction parameter. P*=P/P$_0$, with less than a monolayer formed below P*=0.97. Thicker films form for P*values above the prewetting line, at P*=0.973, 0.979, 0.985, 0.991 and 0.996, from left to right. From Curtarolo et al.[102]*

VII. *Summary and further remarks*

In this paper, we have summarized the status of a number of systems that have undergone wetting transitions or are predicted to exhibit such transitions. Our focus has been the connection between this phenomenon and the gas-surface interactions present in each problem. In the case of wetting transitions of simple gases on alkali metals, the fairly numerous experiments are reasonably consistent with our theoretical understanding of the potential. However, there are large gaps in this picture from both experimental and theoretical perspectives. Hence, there are relatively few cases where one can assess the quantitative adequacy of the potential. This is regrettable because these unusually weak interactions are of great fundamental interest in the general problem of van der Waals forces.

It would be of considerable value to expand both the number of experimental systems and the variety of probes of these systems. Included in our "wish list" for future research efforts would be both thermodynamic and dynamical information about these systems, e.g. measurements of contact angle, atomic-scale friction, density fluctuations and film-spreading near the wetting transition. One important issue involves hysteresis, which is a familiar problem in phase transitions. This was seen in the original experiments showing the wetting transition of $^4$He on Cs and remains to be understood. Almost all of these wetting studies are conducted without using analytical tools, such as scanning tunneling microscopy, for surface characterization. Hence, it is difficult to assess the relative roles of heterogeneous and homogeneous nucleation.

On the theoretical side, a missing piece of the puzzle is the relative absence of fully quantum-mechanical, finite T, studies of the wetting transition of the helium isotopes on Rb and Cs. Such studies would be of particular importance for $^4$He, because these systems represent ideal 2D manifestations of the superfluid transition, without the complication of the poorly understood "inert layer" problem of the onset of superfluidity. While this inert layer is often attributed to heterogeneity, *per se*, there is an observed systematic correlation between the coverage threshold for superfluid onset and the adsorption well-depth, which suggests the importance of intrinsic factors in determining the "inert layer". It should be pointed out that the superfluid transition of $^4$He/Li, for example, is potentially an ideal system because no solid monolayer is expected to occur[103-105].

One of the most general questions of current interest is the relation between wetting transitions on flat surfaces and capillary condensation within porous media. We refer the interested reader to a number of recent publications concerning this subject, including one by Saam in this volume[21,106,107].

We would like to both acknowledge support of this research by NSF and thank the following colleagues and collaborators for many helpful discussions, as well as assistance with figures: Francesco Ancilotto, Mary J. Bojan, Massimo Boninsegni, Carlo Carraro, Moses Chan, Andrew Chizmeshya,  Stefano Curtarolo, Renee Diehl, Rafael Garcia, Susana Hernández, George Hess, Karl Johnson, Kevin Lehmann, Hye-Young Kim, Scott Milner, Mike Pettersen, Will Saam, George Stan, Bill Steele, Peter Taborek, Flavio Toigo, Gianfranco Vidali, Makoto Yao and Xiongce Zhao.

# REFERENCES

1. P. G. de Gennes, "Wetting-statics and dynamics", Rev. Mod. Phys. **57**, 827 (1985)

2. D. Bonn and D. Ross, "Wetting transitions", Rep. Prog. Phys. **64**, 1085-1163 (2001); B. M. Law, "Wetting, adsorption and surface critical phenomena", Prog. Surf. Sci. **66**, 159-216 (2001)

3. J. W. Cahn, "Critical-point wetting", J. Chem. Phys. **66**, 3667-3672 (1977)

4. C. Ebner and W. F. Saam, "New phase-transition phenomena in thin argon films", Phys. Rev. Lett. **38**, 1486-1489  (1977)

5. J. E. Rutledge and P. Taborek, "Prewetting phase-diagram of He-4 on cesium", Phys. Rev. Lett. **69**, 937-940 (1992); J. A. Phillips, D. Ross, P. Taborek and J. E. Rutledge, "Superfluid onset and prewetting of He-4 on rubidium", Phys. Rev. B**58**, 3361-3370 (1998)

6. M. Yao and Y. Ohmasa, "Wetting phenomena for mercury on sapphire", J. Phys.-Cond. Matt. **13**, R297-319 (2001)

7. P. J. Nacher and J. Dupont-Roc, "Experimental evidence for nonwetting with superfluid helium", Phys. Rev. Lett. **67**, 2966-2999 (1991)

8. E. Cheng, M. W. Cole and W. F. Saam and J. Treiner, "Helium Prewetting and Nonwetting on Weak-binding Substrates", Phys. Rev. Lett. **67**, 1007-1010 (1991) and "Phase transitions in multilayer He films", Phys. Rev. B**46**, 13967-13981 (1992); Erratum B**47**, 14661 (1993)

9. E. Cheng, G. Mistura, H. C. Lee, M. H. W. Chan, M. W. Cole, C. Carraro, W. F. Saam and F. Toigo, "Wetting Transitions of Liquid Hydrogen Films", Phys. Rev. Lett. **70**, 1854-1857 (1993).

10. G. Mistura, H. C. Lee and M. H. W. Chan, "Hydrogen Adsorption on Alkali-Metal Surfaces - Wetting, Prewetting and Triple-Point Wetting", J. Low Temp. Phys. **96**, 221-244 (1994); D. Ross, P. Taborek, and J.E. Rutledge, "Wetting behavior of H2 on cesium", Phys. Rev. B **58**, R4274 (1998); M. Poujade, C. Guthmann and E. Rolley, Europhys. Lett. **58**, 837  (2002)

11. J. Klier and A. F. G. Wyatt, "Nonwetting of liquid He-4 on Rb", Phys. Rev. B**65**, 212504 (2002)

12. G. B. Hess, M. J. Sabatini and M. H. W. Chan, "Nonwetting of cesium by neon near its critical point", Phys. Rev. Lett. **78**, 1739-1742 (1997)

13. F. Hensel and M. Yao, "Wetting phenomena near the bulk critical point of fluid mercury", Ber. Bunsenges. Phys. Chem. **102**, 1798–1802 (1998); F. Hensel and M. Yao, Eur. J. Solid State Inorg. Chem. **34**, 861 (1997); F. Hensel, "The liquid-vapour phase transition in fluid metals", Phil. Trans. Roy. Soc. Series A**356**, 97-115 (1998)

14. V. F. Kozhevnikov, D. I. Arnold, S. P. Naurzakov and M. E. Fisher, "Prewetting transitions in a near-critical metallic vapor", Phys. Rev. Lett. **78**, 1735 (1997)

15. D. Ross, J. A. Phillips, J. E. Rutledge and P. Taborek, "Adsorption of He-3 on cesium", J. Low Temp. Phys. **106**, 81-92 (1997); J. E. Rutledge and P. Taborek, "Adsorption of He-3 on cesium surfaces", ibid **95**, 405-411 (1994)

16. K. S. Ketola, S. Wang, and R. B. Hallock, "Anomalous wetting of helium on cesium", Phys. Rev. Lett. **68**, 201-204 (1992) and "Anomalous wetting of helium on cesium", J. Low Temp. Phys. **89**, 601-604 (1992).

17. S. Herminghaus, J. Vorberg, H. Gau, R. Confrdt, D. Reinelt, H. Ulmer, P. Leiderer and M. Przyrembel, "Hydrogen and helium films as model systems of wetting", Annal. der Physik **6**, 425-447 (1997)

18. V. F. Kozhevnikov, D. I. Arnold, S. P. Naurzakov and M. E. Fisher, "Prewetting phenomena in mercury vapor", Fluid Phase Equilibria **150**,625-632 (1998)

19. G. Mistura and M. H. W. Chan, "Adsorption isotherms of helium on Na and on Rb", Physica B**284**, 135-136 (2000)

20. B. Demolder, N. Bigelow, P. J. Nacher and J. Dupont-Roc, "Wetting properties of liquid-helium on rubidium metal", J. Low Temp. Phys. **98**, 91-113 (1995)

21. W. F. Saam, "Wetting, capillary condensation and more" this issue of J. Low Temp. Phys.

22. P. Taborek, this issue of J. Low Temp. Phys.

23. Francesco Ancilotto, Susana Hernández, Manuel Barranco and M. Pi. "Wetting behavior of He-4 on planar and nanostructured surfaces from Density Functional calculations", this issue of J. Low Temp. Phys.

24. E. Cheng, M. W. Cole, J. Dupont-Roc, W. F. Saam and J. Treiner, "Novel Wetting Behavior in Quantum Films," Reviews of Modern Physics **65**, 557 (1993).

25. J. A. Phillips, P. Taborek  and J. E. Rutledge, "Experimental Survey of Wetting and Superfluid Onset of 4He on Alkali Metal Surfaces", J. Low Temp. Phys. **113**, 829-834 (1998); T. McMillan, J. E. Rutledge, and P. Taborek, "Ellipsometry of Liquid Helium Films on Gold, Cesium and Graphite, J. Low Temp. Phys. **138**, 995-1011 (2005)

26. R. B. Hallock, "Review of Some of the Experimental Evidence for the Novel Wetting of Helium on Alkali Metals", J. Low Temp. Phys. **101**, 31 (1995).

27. M. Rauscher and S. Dietrich, "Wetting phenomena in nanofluidics" , Ann. Rev. Mat. Rsch. **38**, 143-172(2008); M. Barranco, R. Guardiola, S. Hernandez, R. Mayol, J. Navarro and M. Pi,"Helium droplets: an overview", J. Low Temp. Phys. **142**, 1-81 (2006)

28. L.W. Bruch, M.W. Cole and E. Zaremba, "Physical Adsorption: Forces and Phenomena", Dover Publishing (Mineola, NY, 2007); Sec. 4.2

29. E. Hult, H. Rydberg, B. I. Lundqvist and D. C. Langreth, "Unified treatment of asymptotic van der Waals interactions", Phys. Rev. B **59**, 4708 (1999)

30. The alert reader will note that this slope is actually negatively infinite (!) as long as T$\neq$ 0, but at low T the argument is still qualitatively valid.

31. E. Cheng, M. W. Cole, W. F. Saam and J. Treiner, "Wetting Temperature Shift of Helium on a Layered Substrate", J. Low Temp. Phys. **89**, 739-742 (1992)

32. P. Taborek and J. E. Rutledge, "Tuning the wetting transition- prewetting and superfluidity of He-4 on thin cesium substrates", Phys. Rev. Lett. **71**, 263-266 (1993)

33. W. Shi, J. K. Johnson and M. W. Cole, "Wetting transitions of hydrogen and deuterium on the surface of alkali metals", Phys. Rev. B**68**, 125401 (2003)

34. S. Curtarolo, G. Stan, M. W. Cole, M. J. Bojan, and W. A. Steele, "Computer simulations of the wetting properties of neon on heterogeneous surfaces", Phys. Rev. E **59**, 4402-4407 (1999)

35. S. Dietrich, "Critical wetting of surfaces with long-range forces", Phys. Rev. B**31**, 4718 (1985)

36. V. B. Shenoy and W. F. Saam, "Continuous wetting transitions in Xe adsorbed on NaF and on plated Cs and Rb substrates", Phys. Rev. Lett. **75**, 4086-4089 (1995)

37. S. Rafai, D. Bonn, E. Bertrand, J. Meunier, V. C. Weiss and J. O. Indekeu, "Long-range critical wetting: Observation of a critical end point", Phys. Rev. Lett.**92**, 245701 (2004)

38. Chizmeshya, M. W. Cole and E. Zaremba, "Weak binding potentials and wetting transitions", J. Low Temp. Phys. **110**, 677-684 (1998)

39. E. Cheng, M. W. Cole, W. F. Saam and J. Treiner, "Wetting transitions of classical liquid films: a nearly universal trend," Phys. Rev. B**48**, 18214-18221 (1993).

40. M. S. Sellers and and J. R. Errington, "Influence of Substrate Strength on Wetting Behavior", J. Phys. Chem. C**112**, 12905-12913 (2008)

41. S. Curtarolo, G. Stan, M. J. Bojan, M. W. Cole, and W. A. Steele, "Threshold for wetting at the triple point", Phys. Rev. E **61**, 1670-1675 (2000)

42. B. E. Clements, H. Forbert, E. Krotscheck and M. Saarela, "He-4 on weakly attractive substrates-structure, stability and wetting behavior", J. Low Temp. Phys. **95**, 849-881 (1994)

43. M. Boninsegni, M. W. Cole and F. Toigo, "Helium adsorption on a Lithium substrate", Phys. Rev. Lett., **83**, 2002-2005 (1999)

44. M. W. Cole and E. Susana Hernández, "Unified model of wetting and pore-filling", Phys. Rev. B**75**, 205421 (2007)

45. Francesco Ancilotto and Flavio Toigo,"Prewetting transitions of Ar and Ne on alkali metal surfaces", Phys. Rev. B**60**, 9019-9025 (1999)

46. J. E. Finn and P. A. Monson, "Prewetting at a fluid-solid interface via Monte Carlo simulation", Phys. Rev. A**39**, 6402-6406 (1989); Y. Fan, J. E. Finn, and P. A. Monson, J. Chem. Phys. **99**, 8238 (1993).

47. V. Apaja and E. Krotscheck, "A microscopic view of confined quantum liquids", in "Microscopic Approaches to Quantum Liquids in Confined Geometries", E. Krotscheck and J. Navarro, editors, pages 205–268 (World Scientific, Singapore, 2002)

48. Francesco Ancilotto, Stefano Curtarolo, Flavio Toigo and Milton W. Cole, "Evidence Concerning Drying Behavior of Ne at the Cs Surface", Phys. Rev. Lett. **87**, 206103 (2001)

49. M. Barranco, M. Guilleumas, E. S. Hernández, R. Mayol, M. Pi and L. Szybisz, "From nonwetting to prewetting: The asymptotic behavior of He-4 drops on alkali substrates", Phys. Rev. B**68**, 024515 (2003)

50. M. J. Bojan, G. Stan, S. Curtarolo, W. A. Steele, and M. W. Cole, "Wetting transitions of Ne", Phys. Rev E **59**, 864-873 (1999)

51. Garcia, K. Osborne and E. Subashi, "Validity of the `sharp-kink approximation' for water and other fluids", J. Phys. Chem. B**112**, 8114-8119 (2008)

52. G. Vidali, G. Ihm, H. Y. Kim and M. W. Cole, "Potentials of Physical Adsorption", Surf. Sci. Repts. **12**, 133-182 (1991)

53. Chizmeshya and E. Zaremba,,"The interaction of rare-gas atoms with metal-surfaces= a scattering-theory approach", Surf. Sci. **268**, 432 (1992)

54. Chizmeshya and E. Zaremba,, "Interaction of rare-gas atoms with metal-surfaces- a pseudopotential approach", Surf. Sci. **220**, 443 (1989)

55. J. B. Anderson, C. A. Traynor and B. M. Boghosian, "An exact quantum Monte-Carlo calculation of the helium helium intermolecular potential", J. Chem. Phys. 99, 345-351 (1993); J. B. Anderson, "An exact quantum Monte Carlo calculation of the helium-helium intermolecular potential.II, J. Chem. Phys. **115**, 4546-4548 (2001)

56. Fig. 6 of Cheng et al (Ref. 39) implies strong wetting for H2/Mg, since D*=5.6 and C3*=4.4

57. Physics Vade Mecum, ed. H. L. Anderson (American Inst. Physics,1981, New York), p. 309

58. CRC handbook on Chemistry and Physics version 2008, p. 112-114.

59. F. Toigo and M. W. Cole, "Model Adsorption Potentials of He and Ne on Graphite," Phys. Rev. B**32**, 6989-6992 (1985); Erratum B**33**, 4330 (1986).

60. M. J. Stott and E. Zaremba, "Quasiatoms- an approach to atoms in nonuniform electronic systems", Phys. Rev. B**22**, 1564-1583 (1980)

61. J. P. Cowin, C.-F. Yu, L. Wharton, Surf. Sci. **161**, 221(1985)

62. U. Harten, J. P. Toennies, C. Wöll, J. Chem. Phys. **85**, 2249 (1986)

63. Antonio Šiber, Ch. Boas, M. W. Cole and Christof Wöll,  "Anomalously low probabilities for rotational excitation in HD/surface scattering: a sensitive and direct test of the potential between closed shell molecules and alkali metals", ChemPhysChem.**7**, 1015-1018 (2006)

64. U. Kleinekathöfer, Chem. Phys. Lett. **324**, 403–410  (2000); D. J. Funk, W. H. Breckenridge, J. Simons, G. Chałasinski, J. Chem. Phys. **91**, 1114 (1989)

65. K. T. Tang, J. P. Toennies and C. L. Yiu, Phys. Rev. Lett. **74**, 546 (1995)

66. James H. Reho, "Time resolved spectroscopy of atomic and molecular dopants in and on helium nanodroplets", Ph. D. thesis, Princeton University, unpublished, 2000

67. J. Pascale, Phys. Rev. A**28**, 632 (1983)

68. P. Jankowski and B. Jeziorski, J. Chem. Phys. **111**, 1857 (1999)

69. Bhattacharya and J. B. Anderson, "An Exact Quantum Monte Carlo  Calculation of the H-He Interaction Potential", Phys. Rev. A **49**, 2441 (1994)

70. Hye-Young Kim, S. M. Gatica and M. W. Cole, "Interaction thresholds for adsorption of quantum gases on planar surfaces, within slit or cylindrical pores and within cylindrical tubes", J. Phys. Chem. A **111**, 12439-12446 (2007)

71. X.-Z. Ni and L. W. Bruch, "Hartree and Jastrow approximations for monolayer solids of Ne, D-2, He-4 and He-3", Phys. Rev. B**33**, 4584 (1986)

72. P. A. Whitlock, G. V. Chester and M. H. Kalos, "Monte-Carlo study of He-4 in 2 dimensions", Phys. Rev. B**38**, 2418-2425 (1988)

73. E. Vitali, M. Rossi, F. Tramonto, D. E. Galli and L. Reatto, "Path-integral ground-state Monte Carlo study of two-dimensional solid He-4", Phys. Rev. B**77**, 180505 (2008)

74. E. Van Cleve, P. Taborek and J. E. Rutledge, "Helium Adsorption on Lithium Substrates", J. Low Temp. Phys. **150**, 1–11 (2008)

75. D. M. Ceperley, "Path-integrals in the theory of condensed helium", Rev. Mod. Phys. **67**, 279-355 (1995)

76. Demolder, F. Raad and J. Dupont-Roc, J. Low Temp. Phys. **101**, 337 (1995).

77. Demolder and J. Dupont-Roc, "Wetting transitions of liquid helium on oxidized rubidium metal surfaces", J. Low Temp. Phys. **104**, 359 (1996)

78. M. S. Pettersen and W. F. Saam, "Prediction of reentrant wetting of He-3-He-4 mixtures on cesium", J. Low Temp. Phys. **90**, 159 (1993) and "Wetting of 3He-4He mixtures on Cesium and Other Alkali Metals," M. S. Pettersen and W. F. Saam, Phys. Rev.B**51**, 15369 (1995).

79. W. F. Saam and M. S. Pettersen, "Wetting of 3He-4He mixtures on Alkali Metal Substrates," J. Low Temp. Physics **101**, 355 (1995) and "Wetting Phenomena in 3He-4He Mixtures on Weak and Superweak Substrates", J. Low Temp. Phys. **110**, 697 (1998).

80. K. S. Ketola and R. B. Hallock, J. Low Temp. Physics **93**, 935 (1993)

81. K. S. Ketola, T. A. Moreau and R. B. Hallock, "Novel behavior of 3He-4He mixture films on a cesium coated substrate at low temperatures", J. Low Temp. Physics **101**, 343-348 (1995)

82. J. E. Rutledge, D. Ross and P. Taborek, J. Low Temp. Physics **101**, 217 (1995).

83. Ross, J.E. Rutledge and P. Taborek,"Wetting transitions of binary liquid mixtures at a weakly attractive substrate, Fluid Phase Equilibria 150–151 , 599-605 (1998)

84. K. R. Atkins, "The surface tension of liquid helium", Canadian J. Phys. **31**, 1165-1169 (1953)

85. M. S. Pettersen, E. Rolley, C. Guthmann and M. Poujade ,"Wetting in Binary Fluid Mixtures: Recent Results in H2/He on Cesium", J. Low Temp. Phys. **134**, 281 (2004).

86. C. Ebner and W. F. Saam, Phys. Rev. B**35**, 1822 (1987)

87. G. Mistura, F. Ancilotto, L. Bruschi and F. Toigo, "Wetting of argon on CO2", Phys. Rev. Lett. **82**, 795-798 (1999)

88. Ancilotto and F. Toigo, "First-order wetting transitions of neon on solid CO2 from density functional calculations", J. Chem. Phys. **112**, 4768-4772 (2000)

89. L. Bruschi, E. Paniz, G. Mistura and G. Galilei, "Triple-point wetting of Ne on solid CO2", J. Chem. Phys. **114**, 1350-1354 (2001)

90. L. J. Munro, J. K. Johnson, and K. D. Jordan, "An interatomic potential for mercury dimer", Journal of Chemical Physics **114**, 5545-5551 (2001)

91. M. E. Fisher, private communication

92. W. Adamson and A. P. Gast, Physical Chemistry of Surfaces, 6th Ed. (John Wiley, NY,1997)

93. J. Morcos, "Surface tension of stress-annealed pyrolitic graphite", J. Chem. Phys. **57**, 1801 (1972)

94. M. E. Schrader, "Ultrahigh-vacuum techniques in the measurement of contact angles. 5. LEED study of the effect of structure on thewettability of graphite", J. Phys. Chem. **84**, 2774 (1980)

95. T. Werder, J. H. Walter, R. L. Jaffe, T. Halicioglu and P. Koumoutsakos, "On the water-carbon interaction for use in molecular dynamics simulations of graphite and carbon nanotubes", J. Phys. Chem. B**107**, 1345 (2003)

96. Pertsin and M. Grunze, "Water-graphite interaction and behavior of water near the graphite surface ", J. Phys. Chem. B**108**, 1357 (2004); K. Karapetian and K. D. Jordan, in Water in Confining Geometries, edited by V. Buch and J. P. Devlin (Springer, Berlin, 2003), pp. 139-150; Xiongce Zhao and J. Karl Johnson, "An effective potential for adsorption of polar molecules on graphite", Molecular Simulation, **31** 1-10 (2005)

97. R. L. Jaffe, P. Gonnet, T. Werder, J. H. Walther and P. Koumoutdakos, "Water–Carbon Interactions 2: Calibration of Potentials using Contact Angle Data for Different Interaction Models", Molecular Simulation **30**, 205 (2004)

98. S. M. Gatica, Xiongce Zhao, J. K. Johnson and M. W. Cole, "Wetting transition of water on graphite and other surfaces", J. Phys. Chem. B**108**, 11704-11708 (2004)

99. X. Zhao, "Wetting transition of water on graphite: Monte Carlo simulations", Phys. Rev. B**76**, 041402 (2007)

100. A. Vogler, "How water wets biomaterial surfaces", in Water in biomaterials surface science", ed. M. Morra (Wiley, NY, 2001), pp. 269-290

101. Rafael Garcia, private communication

102. Stefano Curtarolo, Milton W. Cole  and Renee D. Diehl,"Wetting transition behavior of Xe on Cs and Cs/graphite", Phys. Rev. B **70**, 115403 (2004)

103. G. A. Csathy, J. D. Reppy and M. H. W. Chan, "Substrate-Tuned Boson Localization in Superfluid 4He Films"Phys. Rev. Lett. **91**, 235301 (2003)

104. P. J. Shirron and J.M. Mochel, Phys. Rev. Lett. **67,** 1118 (1991); M.-T. Chen, J.M. Roesler, and J.M. Mochel, J. Low Temp. Phys. **89**, 125 (1992)

105. P.W. Adams and V. Pant, Phys. Rev. Lett. **68**, 2350 (1992)

106. Hye-Young Kim, S. M. Gatica and M. W. Cole, "Interaction thresholds for adsorption of quantum gases on planar surfaces, within slit or cylindrical pores and within cylindrical tubes", J. Phys. Chem. A **111**, 12439-12446 (2007).

107. Milton W. Cole and E. Susana Hernández, "Unified model of wetting and pore-filling", Phys. Rev. B**75**, 205421 (2007)